\documentclass[useAMS,usenatbib]{mn2e}

\usepackage{graphicx}

\title[Clump formation due to the gravitational instability of a multiphase medium]{Clump formation due to the gravitational instability of a multiphase medium in a massive protoplanetary disc}
\author[V. N. Snytnikov and O. P. Stoyanovskaya]{V. N. Snytnikov\thanks{E-mail:
snyt@catalysis.ru (VNS); stop@catalysis.ru (OPS) } and O. P.
Stoyanovskaya \\
Boreskov Institute of Catalysis SB RAS, Novosibirsk, Lavrentieva,
5, 630090, Russia}

\begin{document}

\date{Accepted ***. Received ***; in original form ***}

\pagerange{\pageref{firstpage}--\pageref{lastpage}} \pubyear{2002}

\maketitle

\label{firstpage}

\begin{abstract}

Planetary systems form in gas-dust protoplanetary discs via the
growth of solid bodies. In this paper, we show that the most
intriguing stage of such growth --- namely, the transformation of
1-10 m boulders into kilometre-sized planetesimals --- can be
explained by a mechanism of gravitational instability. The present
work focused on the origin of self-gravitating clumps in which
planetesimal formation could take place. Our computer simulations
demonstrated that such clumps of gas and boulders formed due to
the development of a two-phase instability. This instability
revealed a so-called 'mutual influence effect' in the
protoplanetary disc, where the dynamics of the system were
determined by the collisionless collective motion of a low-mass
subdisc composed of primary solids. We found that a 0.1 $c_s$
velocity dispersion in the boulder subdisc was sufficient to cause
the formation of self-gravitating clumps of gas and boulders. In
such regimes, the time needed for the formation of the collapsing
objects was less than the boulders' dissipation time in the
density waves of the medium.
\end{abstract}


\section{Introduction}

Results from numerical modelling and observations of the gas-dust
discs around young solar-type stars provide the basis for the
currently accepted scenario for the formation of the Solar System (Hartmann 2009).
This model suggests that the disc forms simultaneously with a
protostar, following the gravitational collapse of gas in the
molecular cloud (Petit \& Morbidelli 2005). In the first stage
(which lasts up to 100,000 years), a protostar forms; this
consists mainly of hydrogen and helium, and has a mass of about
one tenth of the solar mass.

The massive gas-dust disc is formed by the collision of the
opposing gas streams. The gas streaming during the molecular cloud
collapse is supersonic (Spitzer 1978). The infalling gas streams
collide and produce the diverging shock waves (Landau \& Livshitz
1987)that decelerate the gas streams velocity (see Fig. 1a).
Between a pair of shock waves the  gas density is higher than its
extreme value for a single shock wave, as it was shown
theoretically and experimentally by Orishich, Ponomarenko \&
Snytnikov 1989. When the flow rate of collapsing gas decreases in
the end of the star formation stage, the shock waves diverge from
the disc plane (see Fig. 1b). Now the gas can leave the disc. Fast
diminution of its flow provides conditions for the fast gas
expansion which is followed up with its cooling.

\begin{figure*}
  \includegraphics[width=150mm]{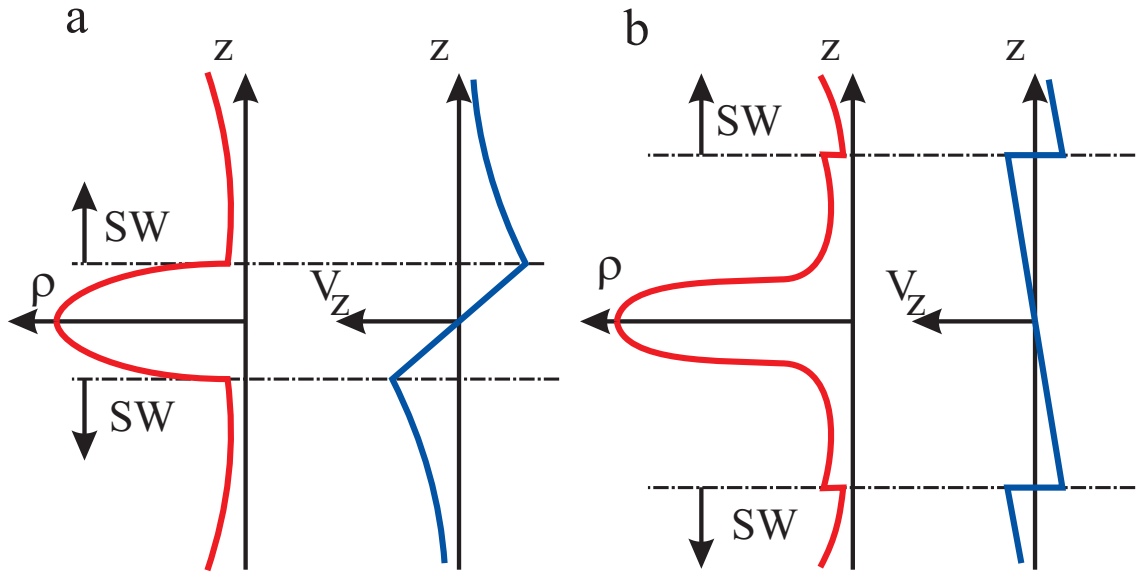}
  \caption{: The formation of the massive gas-dust disc by the collision of the opposing streams during the molecular cloud collapse. }
\end{figure*}

The molecular cloud dust moves together with the gas. During the
gas compression behind the shock waves (Fig. 1a) the dust grains
grow in size as a result of heavy molecular absorption and
coagulation (Cuzzi \& Weidenschilling 2006). Adsorption time is
less than $10^6$ years in a medium with the hydrogen concentration
higher than $10^4$ cm$^{-3}$ (Spitzer 1978). In the disc the
concentration of gas is by several orders of magnitude higher than
$10^4$ cm$^{-3}$. So the molecules are adsorbed according to their
sublimation temperature. Spitzer (1978) estimated that the
coagulation length exceed the absorption length in the order of
1-2. These lengths are significantly less than the massive disc
thickness. That is why the dust grains can grow in size in
accretion disc, settle on the disc plane and form their own
subdisc (Cuzzi \& Weidenschilling 2006). Due to the subdisc gas
expansion in outside flow the relation of solid density to gas
density increases in contrast to the molecular clouds, where dust
constitutes 1-2 per cent of the mass.

The protostar increases its mass up to the star mass due to
accretion from the disc and the residues of the surrounding
molecular cloud. Alteration of the disc mass is defined by the
flows: (a) falling to the disc from the cloud, (b) accreting from
the disc to the protostar, (c) leaving the disc, e.g. in jets. By
the time discs become observable (1-3 million years), their masses
have decreased to 1 per cent of the host star mass. Inside the
period between protostar and young star a stage of massive disc
exists, when the mass of the disc and the mass of the central body
are comparable. Over the course of the next 60-100 million years,
the circumsolar disc evolves to a state similar to that currently
present in our own Solar System.

The formation of planets in circumstellar discs proceeds via the
growth of solids; nanometer-sized dust in molecular clouds
transforms into planets with radii of some thousands of
kilometres.

The growth of dust grains via coagulation  depends on chemical
composition. Usually ice, silicates and iron are mentioned as dust
compounds. But due to the cosmic abundance of chemical elements
(Lodders \& Fegley 1998) the main components of disc solids dust
should be organic compounds from H, C, O, N with high molecular
weights (Herbst \& van Dishoeck 2009). Such compounds can be
easily synthesized on Mg, Si, O, Fe dust grains that work as
catalysts in a medium with abundance of H and He (Khassin \&
Snytnikov 2005). That is why conglutination of multi-component
organic compounds and non-organic bodies can be expected up to
1-10 meter in size. When the subdisc consists of such bodies with
velocities greater than 10 km s$^{-1}$ they collide less than once
per orbital time.

The growth of planetesimals (i.e., bodies with a size in excess of
a kilometre) may occur due to collisional accumulation, since
their own gravitational field can attract smaller bodies and hold
these to their surface (Safronov 1969). The mechanism by which
kilometre-sized planetesimals are formed from 'boulders' (or
agglomerates, with sizes from $1-10$ m) has been the subject of
intense interest and discussion, and has formed the basis of many
studies (Youdin \& Shu 2002, Makalkin \& Ziglina 2004, Rice et al.
2006, Johansen et al. 2007). At this stage, the concentration of
solids provides motion, with rare collisions at orbital time with
velocities of tens kilometers per second. The metre-sized boulders
are not enlarged during high-velocity collisions. Collisions of
such boulders don't result in sticking irrespective of their
composition. During collisions where the relative velocities
exceed $1$ m s$^{-1}$, inorganic compound solids larger than 10 cm
are destroyed, rather than sticking together and becoming a
larger, aggregate body (Marov et al. 2008). In addition, in
circumstellar discs, the time taken for a metre-sized solid to
fall to the central body is of the order of 100 years, due to gas
drag. Thus, the growth of such solids in the disc may take only
some tens of rotations (Wiedenschilling 2000, Armitage 2006). If
the growth goes slower, the protoplanetary disc will lose a large
proportion of its 'solid' matter, which is the material needed for
the formation of Earth-group planets, and asteroids and comets.

The formation of planetesimals can occur either in a massive
circumstellar disc (age 0.1-1 million years, with the mass of the
disc comparable to that of the central body) or in a medium-mass
disc (age exceeding 1 million years, with the mass of the disc an
order of magnitude less than the mass of the central body).

One possible way to rapidly assemble metre-sized boulders into a
body of planetismal size could be via gravitational instabilities
triggered by the collective motion of the solids subdisc.

Gravitational instability in discs with a central body has been
investigated since the 1950s (Edgeworth 1949). It was found that
to experience local Jeans instability, a disc with a central body
needs lower dispersions in the velocities of solids (Gurevich \&
Lebedinsky, 1950), as well as lower gas temperatures (Safronov
1960) or higher-density matter (as compared with a motionless
medium). Such conditions --- which are close to those that trigger
other instabilities in the disc (Fridman 2008) --- may lead to the
formation of rings, spirals, complex wave structures or individual
clumps (Snytnikov et al. 2004). The formation of clumps in a
rotating medium is hindered by gas heating, as well as the
increasing velocity dispersion of solids produced by gravitational
field perturbations in spiral and other wave structures. The
non-linear behaviour of instabilities and the appearance of clumps
(whose densities increase under self-gravitation) are therefore of
interest. In any case, for discs to experience fragmentation, they
should be quite dense and cold. Scenarios for the development of
instabilities (Levin 1965, Goldreich \& Ward 1973) suggest that
stable discs transform into unstable ones due to the sedimentation
of solids on the equatorial plane. For instability development it
was necessary to provide the surface density of the primary solids
critical for this particle velocity dispersion. However, the
formation of a dense primary solids subdisc is prevented by solids
scattering, which is caused by turbulent gas flows resulting from
hydrodynamic instability of the Kelvin-Helmholtz type (Cuzzi,
Dobrovolskis \& Champney 1993). This instability arises because
both components of the two-phase subdisc located in the equatorial
plane rotate at the Keplerian velocity, whereas the angular
velocity of gas outside the plane is lower, due to the radial
pressure gradient. A substantial difference in subdisc thickness
(with respect to the gas and condensed phases) is not therefore
expected.

On the other hand, a disc can reach its instability threshold not
only by increasing its density, but also via a decrease in
temperature. The main problem with implementing this mechanism in
a medium-mass disc is that the lowering of the gas temperature in
the disc is accompanied by a rapid formation of spiral structures,
which do not transform into clumps. According to estimates made by
Makalkin \& Ziglina (2004) and Marov et al. (2008), at the stage
of planetesimal formation in the Solar System, the temperature of
the gas component in the disc at the Earth's radius was 500-700 K
when the disc had uniformly distributed solids, and 250-500 K
after the formation of a
thin subdisc of primary solids. Spiral waves appear at higher
temperatures in the medium-mass disc. In these spirals, the
dispersion in the velocities of solids increases (Rice et al.
2004). These structures --- which result from the development of
gravitational instability in the gas --- scatter the subdisc of
primary solids with sizes exceeding 1 m (Rice et al. 2006). Rapid
cooling, which typically occurs in a time comparable to the time
taken for the disc rotation, was considered to be a necessary
condition for clump formation in such a disc (Gammie 2001, Rice et
al. 2003). This imposes stringent restrictions on the mass and
temperature of the disc, as well as on the zones of formation of
such clumps (Rafikov 2009). Further studies revealed that such
restrictions could not provide conclusive evidence, proving --- or
disproving --- that gravitational instability is the mechanism
responsible for the formation of large bodies in the discs. The
reason is that the critical time of cooling depends on the
$\gamma$ - constant ratio of the specific heats (Rice et al.
2005), the history and rate of cooling (Clarke, Harper-Clark \&
Lodato 2007), and the ratio of the mass of the disk to the mass of
the central body (Meru \& Bate 2010). Thus, for a disc whose mass
is 10 times smaller than that of the central body, a cooling time
equal to 15 disc rotations is now taken to be critical (Lodato \&
Clarke 2011). However, in the case of a medium-mass disc, such gas
cooling conditions can hardly be expected to occur, due to the
growing radiation from the star.

The mechanisms that facilitate the formation of planetesimals from
large bodies have previously been studied for medium-mass discs.
According to the computer simulation made by Rice et al. (2004,
2006) for a two-phase disc, metre-sized solid particles can
concentrate in spiral arms for some time, due to gas dragging.
Computational experiments by Youdin \& Shu (2002) demonstrated
that a turbulent gas flow resulting from the difference in the
angular velocities of the primary solids subdisc and the gas disc
did not prevent gravitational instability which increases the
concentration of solids. A condition for the development of
instability is determined by the compression of the solid phase
subdisc towards the star, in the equatorial plane. Such
compression increases the ratio of the mass densities of the solid
component and the gas by a factor of 2-10. The density of the
medium in these regions starts to grow under the action of a
self-gravitational field. Marov et al. (2008) noted that the
possibility of a local concentration of bodies should be
considered; this local concentration may result from the
differential rotation of the gas subdisc in large turbulent
structures. According to Cuzzi et al.(1993), such bodies will be
represented by metre-sized boulders, since solids exceeding 1 m in
size settle most efficiently on the equatorial plane. Meanwhile,
smaller agglomerates are removed from this plane by a turbulent
flow that results from the different angular velocities of the gas
and solid-phase subdiscs.

Overall, the problem of planetesimal formation from primary
solids, boulders and agglomerates in medium-mass quasi-stationary
discs has not yet been convincingly solved. We therefore examined
the possibility of planetesimal formation in the intermediate
period between the existence of the massive accretion disc (age of
more than 0.1 million years, with the mass of the disc comparable
to the mass of the central body) and the medium-mass disc (age of
more than 1 million years, with the mass of the disc approximately
10 times smaller than the mass of the central body).

For massive disc formation Machida, Inutsuka \& Matsumoto (2008)
simulated the structure of the gas flow and demonstrated that the
streams outside the disc plane along the rotation axis must exist.
These outgoing flows can provide intense non-radiative cooling of
gaseous disc under the condition of weak radiative heating from
protostar of a small mass.

Johansen et al. (2007) investigated local gravitational collapses
in medium of sub-metre-sized solids in the disc. In their
simulations the dispersion in the velocities of such solids was
determined by turbulent fluctuations in the gas flow. Owing to the
primarily collective motion of the gas and solids, the dispersion
may be as high as ten metres per second. Johansen et al. (2007)
found that in a local region with a self-gravitating medium and an
external gravitational field exerted by a protostar, the drag
force resulted in gravitational collapse in a subsystem of solids,
over several orbital periods. Gas was not involved in such
collapses. This mechanism was able to explain the growth of
sub-metre-sized agglomerates into bodies with a size of 10 metres.
These 10-metre-sized bodies experience rare but high-velocity
collisions with each other during orbital time. In contrast to
sub-metre-sized solids, drag forces do not cause such enlarged
bodies to move in concert with gases. The collective motion of
these bodies in the disc is described by the Vlasov equation.

The main thesis of this paper is that the gravitational
interaction of the two phases (gases and primary solids over a
metre in size) affects the stability of the entire disc, and
changes the conditions necessary for the formation of clumps in
such a system. We investigated the gravitational interactions
between the gas and solid phases in a massive disc, during the
development of an instability that could lead to the formation of
clumps. As it was explained above the surface density of the solid
phase was assumed to be greater than 1 per cent of the gas surface
density. The solids subdisc reaches its fragmentation threshold
only due to presence of  massive gaseous envelope.

The second section of the paper describes a mathematical model for
the two-phase protoplanetary disc during clump formation, and the
third section deals with the numerical algorithms and code we used
in the computational experiments. The fourth section presents
estimates for the effective Jeans lengths of the gravitational
interactions between the gases and the primary solids subdisc. The
fifth section shows the results of numerical modelling for the
dynamics of the two-phase system with a central body, for time
periods covering several disc rotations.

\section[]{A mathematical model of the protoplanetary disc during clump formation}

\subsection{Basic equations}

The computational experiments reported in this paper were carried
out within a quasi-3D model of the disc. This means that we
neglected the vertical motion of matter, and considered the
dynamics of the razor-thin disc where its entire mass was
concentrated inside the equatorial plane of the system. The
gravitational potential of the \textbf{\textit{multiphase}} disc
was calculated using the 3D Laplace equation.

The limits of such model applied to the disc description were
discussed in detail by Fridman \& Khoruzhii (2003). They found
that the model is valid, first, when the typical length of density
perturbation in the disc is greater than its thickness; secondly,
when the changes of the dynamical processes time are slow with
respect to orbital time. The first condition is satisfied for all
process in the disc except for its collapsing. The second
condition is satisfied for gravitational instability almost
everywhere inside the disc except for its outer edge where
dynamical and orbital time are comparable.

The gas component was described by the following gas dynamics
equations:

\[
\frac{\partial \sigma}{\partial t}+div(\sigma
\textbf{\textit{v}})=0,
\]
\[
\sigma \frac{\partial \textbf{\textit{v}}}{\partial
t}+\sigma(\textbf{\textit{v}},\nabla)\textbf{\textit{v}}=-\nabla
p^* - \sigma\nabla\Phi,
\]
\[
\frac{\partial S^*}{\partial t}+(\textbf{\textit{v}},\nabla) S^* =
0, \ \ \ \ p^*=T^*\sigma. \ \ \
\] These gas dynamics equations include surface quantities that were
obtained from volume quantities by integration with respect to the
vertical coordinate $z$:
\[\sigma=\sigma_{gas}=\int_{-\infty}^{+\infty}\rho_{gas}dz; \ \ \
p^*=\int_{-\infty}^{+\infty} p dz.\]

Here, $\textbf{\textit{v}}=(v_x,v_y)$ is the two-component gas
velocity, and  $p^*$ is the surface gas pressure.
$T^*=\displaystyle\frac{p^*}{\sigma}$, $S^* = \ln
\displaystyle\frac{T^*}{\sigma^{\gamma^*-1}}$ are the quantities
similar to gas temperature and entropy. $\gamma^*$ is a 2D version
of $\gamma$(Polyachenko \& Fridman 1976), which is related to the
constant ratio of specific heats as
$\gamma^*=3-\displaystyle\frac{2}{\gamma}$. $\Phi$ - is the
gravitational potential in which the motion occurs.

In our model, the solid-phase subdisc was represented by 1-10 m
solids moving in such a manner that two solids collided with a
frequency not exceeding one event per rotation around the
protostar. The dynamics of the primary solids subdisc were
described by the Vlasov equation, neglecting collisions between
solids at times longer than several rotations: \[\frac{\partial
f}{\partial t}+\textbf{\textit{u}}\frac{\partial f}{\partial
\textbf{\textit{r}}}+\textbf{\textit{a}}\frac{\partial f}{\partial
\textbf{\textit{u}}}=0, \ \ \ \ \ \] where
$\textbf{\textit{a}}=-\nabla{\Phi},$ $\textbf{\textit{a}}$ is the
acceleration of particles in an external and self-consistent
field, $\textbf{\textit{u}}=(u_r,u_{\phi})$ is the two-component
velocity of the particles, and
$f=f(t,\textbf{\textit{r}},\textbf{\textit{u}})$ is a function of
the particle velocity distribution $\textbf{\textit{u}}$ at a
point in the disc with the coordinate $\textbf{\textit{r}}$. This
function is related to the surface density of the particles as
$\sigma_{par}=\int f d\textbf{\textit{u}} dz.$ Note that the
radius of the solids does not appear explicitly in the Vlasov
equation; however, the description of the primary solids subdisc
as a collisionless system implies a certain size range for these
solids.

$\Phi$ is the gravitational potential, which is the sum of the
potential of the motionless central body and the potential of the
razor-thin disc,  $\Phi=\Phi_1+\Phi_2,
\Phi_1=-\displaystyle\frac{M_c}{r},$ where $M_c$ is the mass of
the central body. $\Phi_2$ is the potential of the self-consistent
gravitational field, which is defined in general from the
Dirichlet problem for the 3D Poisson equation
\[\Delta\Phi_2=4 \pi (\rho_{par}+\rho_{gas}), \ \ \ \Phi_2\longrightarrow_{\sqrt{r^2+z^2}\rightarrow
\infty} 0.\] For
$\rho_{par}+\rho_{gas}=(\sigma_{par}(r,\phi)+\sigma_{gas}(r,\phi))\delta(z)$,
where $\delta(z)$ is the Dirac delta function, $\Phi_2$ is found
as a solution of mixed problem for the Laplace equation
\[\Delta\Phi_2=0, \ \ \ \Phi_2\longrightarrow_{\sqrt{r^2+z^2}\rightarrow \infty} 0,\]
\[\frac{\partial \Phi_2}{\partial z}|_{z=0} = 2 \pi
(\sigma_{par}(r,\phi)+\sigma_{gas}(r,\phi)).\]

The equations are written using dimensionless variables. The basic
dimensional quantities are $G$ (the gravitational constant), and
$R_0=10$ au, $M_0=M_{\odot}=2 \cdot 10^{30}$ kg, which are the
typical size and mass of the system. The gas component and the
boulders subdisc interact through a common gravitational field.

\subsection{Initial conditions}
The surface temperature and density of the disc were specified at
zero time. In the calculations presented in this paper, the
density of the gas and primary solids subdisc was taken as a
Mackloren disc of mass $M_{par,gas}$ and radius $R$:
\[\sigma_{par,gas}(r)=\left\{\begin{array}{l}
  \displaystyle\frac{3M_{par,gas}}{2 \pi R^2}\sqrt{1-(\frac{r}{R})^2}, \ \ r<R,
  \\[5mm]
  0, \ \ r\geq R.
  \end{array} \right.\]

The gas temperature at zero time was specified as $T^*(r) \sim
(1-\displaystyle\frac{r}{R})$ or $T^*(r) \sim \sigma(r)$, using a
given $T_0$, which is the temperature in the 'centre' of the disc.

The initial velocities of the solids were specified as the sum of
regular and random components,
$\textbf{\textit{u}}=\textbf{\textit{u'}}+\textbf{\textit{u''}}$,
where $\textbf{\textit{u'}}$ is the regular velocity, and
$\textbf{\textit{u''}}$ is the random velocity. The gas velocity
and the regular velocity of particles were determined from an
equilibrium between centrifugal and centripetal gravitational
forces:
$\displaystyle\frac{v_{\phi}^2}{r}=\frac{1}{\sigma}\frac{\partial
p^*}{\partial r}+\frac{\partial \Phi}{\partial r},$
$\displaystyle\frac{u_{\phi}^{'2}}{r}=\frac{\partial
\Phi}{\partial r},$ where $v_r=0,$ and $u_r^{'}=0$. The random
velocity of the particles $\textbf{\textit{u''}}$ was specified by
the Gaussian law with a zero average and a prescribed dispersion
$v_d$.

\section[]{Numerical methods and setups}

We carried out our computer simulation of protoplanetary disc
dynamics using a Sombrero code based on the method of splitting
with respect to the physical processes involved (Stoyanovskaya \&
Snytnikov 2010). The Vlasov equation, the gas dynamics equations,
and the mixed problem for the Laplace equation were solved at each
time step.

The solution of the Vlasov equation was obtained using the
particle-in-cell (PIC) method (Hockney 1987). We combined it with
the grid method solution of the Laplace equation. We interpolated
the surface density of solid bodies on the regular polar grid
using the PIC kernel. The equations of each particle motions are
characteristics of the Vlasov equation: $
\displaystyle\frac{d\overrightarrow{u}}{dt}=-\nabla\Phi, \ \
\frac{d\overrightarrow{r}}{dt}=\overrightarrow{u}.$ They were
integrated by leap-frog scheme (Snytnikov et al. 2004).

The gas dynamics equations were solved using the SPH method
(Monaghan 1992). The SPH calculation formulas implemented in
Sombrero were obtained from quasi-3D gas dynamics equations,
written in the Lagrangian form:
\[ \frac{d
\sigma}{dt}=\sigma \cdot \textrm{div}{\textbf{v}}, \ \
\frac{d\textbf{v}}{dt}=-\frac{1}{\sigma} \nabla p - \nabla \Phi, \
\ \frac{d\textbf{r}}{dt}=\textbf{v}, \ \ \frac{dS}{dt}=0,
\] where $\displaystyle\frac{d}{dt}=\frac{\partial}{\partial t}+ v \cdot
\nabla$.

We used the cubic spline for 2D space as a kernel $W$:
\[
W(q,h)=\frac{5}{14 \pi h^2}\left\{ \begin{array}{l}
                 [(2-q)^3-4(1-q)^3],  \ \ 0\leq q \leq 1,\\
                 \ \ \ \ \ \ [2-q]^3,  \ \ \ \ \ \ \ \ \ \ \ \ \ \ 1 \leq q \leq 2, \\
                 0, \ \ \ \ \ \ \ \ \ \ \ \ \ \ \ \ \ \ \ \ \ \ \ \ \ \ \ \ \ \  \ \ \ q > 2,
                 \end{array} \right.
\]
where $q=\displaystyle\frac{|\textbf{x}|}{h},$ $\textbf{x}$ -
radius vector of a space point.

The surface density of the gas where the \textit{i}th particle
resides was calculated as the sum interpolant $\sigma_i =
\sum_{j=1}^{N} m_j W_{ij},$ where $N$ was the number of simulated
SPH particles. We used the adaptive smoothing length
$h_i=\displaystyle 2\sqrt{\frac{m_i}{10^{-3}+\sigma_i}},$ which
was implemented through arithmetic kernel averaging
\[W_{ij}=\frac{1}{2}\left(W(|r_i-r_j|,h_i)+W(|r_i-r_j|,h_j)\right).\]

The equation of motion was approximated such that the impulse and
angular momentum were preserved:
\[
\frac{d \textbf{v}_i}{dt}=- \sum_j
m_j(\frac{p_j}{\sigma_j^2}+\frac{p_i}{\sigma_i^2}+\Pi_{ij})
\nabla_i W_{ij} -\nabla \Phi_i, \]

\[ \Pi_{ij}=\left\{ \begin{array}{l}
                 \displaystyle\frac{-\alpha \overline{c_{ij}} \mu_{ij}+\beta \mu_{ij}^2}{\overline{\rho_{ij}}},  \ \ v_{ij}r_{ij}<0,\\
                 0, \ \ \ \ \ \ \ \ \ \ \ \ \ \ \ \ \ \ \ \ \  \ v_{ij}r_{ij}\geq 0,,
                 \end{array} \right. \]
\[
\mu_{ij}=\frac{h_{ij}v_{ij}r_{ij}}{|r_{ij}|^2+0.01h_{ij}^2},
 \ \overline{c_{ij}}=\frac{1}{2}(c_i+c_j), \ \overline{\sigma_{ij}}=\frac{1}{2}(\sigma_i+\sigma_j),
\]
\[
 h_{ij}=\frac{1}{2}(h_i+h_j),
v_{ij}=v_i-v_j, r_{ij}=r_i-r_j.
\] For calculation formulas we used notations:
\[
W_{ij}=W(|r_i-r_j|,h), \ \ \nabla_i
W_{ij}=\displaystyle\frac{x_i-x_j}{h}\frac{\partial
W_{ij}}{\partial q}.
\]

We used a standard artificial viscosity with parameters $\alpha=1,
\ \beta=1$ (Monaghan 1992) in Sombrero code to make the
calculation of shearing motion possible. In our model we used the
entropy equation, which doesn't describe the shock waves. That is
why we introduced an artificial viscosity to the motion equation
only. The kinetic energy of gas decreased due to the viscosity on
10 per cent $100\times 128 \times 100$ grid cells and 40000 SPH
particles for the test calculation. The contribution of artificial
viscosity is decreased by increasing number of SPH particles.
Numerical experiments show that the velocity field in clumps is
not affected by the number of SPH particles. If we change the
model by direct artificial heating it provides significant
increasing of gas temperature and the transition of its flow from
supersonic to subsonic in the inner part of the disc.

The Laplace equation was solved using a cylindrical region of
space, with its lower face determining the disc plane. The
particles simulating the gas disc and the primary solids subdisc
were located in this plane. The Laplace equation was solved for
the entire volume of the cylinder, using a cylindrical coordinate
system. A boundary condition fixing the potential to zero at
infinity was transferred to the side surfaces of the cylinder and
the upper edge of the calculated region at each time step, via the
decomposition of the potential into multifields (up to quadrupole
moments). The radius of the calculated region was typically twice
as large as the initial radius of the disc, and the height of the
cylinder was chosen to be equal to the radius of the calculated
region. The Laplace equation was solved using an iterative
combined method, where a value from the previous time step was
taken as an initial approximation. This method employed fast
Fourier transforms in angular coordinates, combined with a
successive block over-relaxation procedure (Snytnikov et al.
2004).

The gravitational force affecting the SPH and PIC particles was
calculated  by a linear interpolation of the force rate in the
mesh points which was obtained from the \textit{midplain}
gravitational potential value.

Most of the calculations presented below were performed on a $100
\times 128 \times 100$ grid. The particles did not reach the
boundary of the calculated region in any of our calculations. To
verify that the changes observed in the solutions were due to the
parameters for the low-mass component of the disc (and were not
related to numerical effects), all calculations were performed
using the same numerical algorithm parameters. The gas disc was
represented by $4 \times 10^4$ SPH particles. In this case, the
mass of the particles within the smoothing length did not locally
exceed the corresponding Jeans mass (Bate \& Burkert 1997). The
primary solids subdisc was represented by $5 \times 10^6$ PIC
particles. In the disc plane, the grid size was $[r,\phi]=100
\times 128$ with $50 \times 128$ cells representing the disc at
zero time. The average number of PIC particles in each cell in the
disc plane was therefore 1000, which provided a $\leq 3 $ per cent
fluctuation in the density and other calculated quantities.

The resolution requirements for multi-phase and single-component
medium  studies are similar. Usually the Jeans length in gas and
subdisc of solids is greater than a few grid cells and smoothing
lengthes of SPH particles. In the next section we discuss
existence of effective Jeans length for gas-solid bodies medium.
Hereby we suppose that a maximum grid cell size and a smoothing
length must be less than the effective Jeans length.

In addition, the applicability of the methods implemented in
Sombrero for the simulation of the dynamics of axisymmetric and
radial-azimuthal structures (that form in the two-phase medium of
the gravitating disc) was verified; this was achieved by comparing
the results of computational experiments performed using the SPH
and FLIC methods (Stoyanovskaya \& Snytnikov 2010). It was shown
that the SPH method was able to reproduce nonlinear waves in the
gas and collisionless solids medium, with shear and counter flows
emerging in the system.

\section[]{Gravitational instability in a medium of gas and collisionless
primary solids medium: Analytical expectations}

The scale separating the growing and decaying perturbations in a
gravitating medium is determined by the local Jeans length. The
Jeans length is obtained from an analysis of the dispersion
ratios, which gives an estimate of the minimum density
perturbation in a system developing under the action of a
self-gravitational field. The long-wave stability of a
collisionless, flat layer was considered by Fridman \& Polyachenko
(1984). They demonstrated that the critical length of the wave is
equal to
\[\displaystyle\Lambda_{par}=\frac{T_{par}}{G \sigma_{par} m_{par}}=
\frac{v_{d}^2}{G \sigma_{par}}.\] The vibrations of the
incompressible gas layer (in the same long-wave approximation,)
are expressed by the relationship
\[\displaystyle\Lambda_{gas}=\frac{T_{gas}}{G \sigma_{gas}
\mu}=\frac{c_{gas}^2}{G \sigma_{gas}}.\]
Similar expressions for the critical extent of the distortions are
available for the Toomre instability in a heterogeneous rotating
disc. The dynamics of gases and collisionless primary solids under
a common gravitational field (in the same long-wave approximation
of a flat layer in a circumstellar disc, for perturbation of a
potential $\tilde\Phi$ near the equatorial plane, in scale
$\Lambda$) can be written as $\tilde{\Phi} \sim
G(\tilde{\sigma_{par}}+\tilde{\sigma_{gas}}) \Lambda$, where
$\tilde{\sigma_{par}},$ and $\tilde{\sigma_{gas}}$ are
perturbations in the particle and gas surface densities. For
rapidly growing waves (rapid compared with the rotation period
around a protostar), with reference to \[\frac{1}{\mu} T
\frac{\widetilde{\sigma_{gas}}}{\Lambda} \sim \frac{\sigma_{gas}
\tilde{\Phi}}{\Lambda},\ \  v_d^2 \widetilde{\sigma_{par}} \sim
\sigma_{par} \tilde{\Phi},\] one can obtain
$\displaystyle\frac{1}{\Lambda} \sim \frac{1}{\Lambda_{par}} +
\frac{1}{\Lambda_{gas}},$ or
\[\Lambda=\displaystyle\frac{\Lambda_{gas}}{1+\frac{c_s^2}{v_d^2}\displaystyle\frac{\sigma_{par}}{\sigma_{gas}}}=\frac{\Lambda_{par}}{1+\frac{v_d^2}{c_s^2}\frac{\sigma_{gas}}{\sigma_{par}}},\]
where $m_{par}v_d^2$ is a typical 'temperature' for the velocity
distribution function of solids (neglecting motion perpendicular
to the disc plane), $m_{par}$ is the mass of a solid particle, and
$T$ and $\mu$ are the temperature and molecular weight of the gas.
$\Lambda$ gives the critical wavelength for a two-component disc,
the so-called effective Jeans length for a two-phase system.

Relation
$Q=\sqrt{\displaystyle\frac{\Lambda_{Jeans}}{\Lambda_{rot}}}<1$
gives the classical Toomre value for the gravitational instability
in the rotating disc, where $\Lambda_{rot}=\displaystyle\frac{G
\sigma_{gas}}{\Omega^2}$, $\Omega$ - epicyclical frequency. For
multiphase disc $\sigma=\sigma_{par}+\sigma_{gas}$, which defines
$\Lambda_{rot}=\displaystyle\frac{G
(\sigma_{gas}+\sigma_{par})}{\Omega^2}$. Then, using the obtained
value of effective Jeans length the Toomre parameter can be
written as
$Q=\sqrt{\frac{\Lambda_{par}\Lambda_{gas}}{\Lambda_{gas}+\Lambda_{par}}
\displaystyle\frac{\Omega^2}{G (\sigma_{gas}+\sigma_{par})}}.$

Inside the disc, for a mixture of hydrogen and helium at a
temperature $T \approx 300$ K, $c_s \approx 1000$ m s$^{-1}$. For
$\sigma_{par}/\sigma_{gas} \approx 5*10^{-2}$ (a massive gas
disc), and with a dispersion in the solid velocities of $v_d
\approx 100 $ m s$^{-1}$, the length $\Lambda$ decreased 6-fold
compared with $\Lambda_{gas}$; for $v_d \approx 20$ m s$^{-1}$, a
100-fold decrease was observed. It was therefore shown that the
presence of a second low-mass component could strongly decrease
the effective Jeans length for the two-phase system $\Lambda$. By
developing the aforementioned instabilities in its subsystem, the
low-mass phase was able to facilitate the development of
gravitational instability in the entire medium. The role of the
gas was reduced to providing a drag force on individual solids;
such friction decreased the velocity dispersion of the primary
solids. The gas therefore served as a massive medium, in which
density perturbations transferred from the primary solids subdisc
could develop, and produce clumps. The physics of the process is
similar to the cooling of heavy ions by light electrons (which
move in concert with the heavy ions) that occurs in accelerator
physics (Nikitin, Snytnikov \& Vshivkov 2004).

\section[]{Computational experiments}

We will now clarify how the gravitational interactions between the
cloud of low-mass collisionless primary solids and the massive gas
can affect the formation and development of structures in the
self-gravitating disc, on its rotation around the central body.
Three cases can be distinguished: first, structures can form and
develop due to instabilities in the gas, and can subsequently
involve primary solids in their dynamics. Second, structures can
be initiated in the primary solids subdisc, and can remain there
for their entire lifetime, without any noticeable effect on the
gas. Third, structures can be generated in the gas, but under the
development of 'two-phase' instability, where the overall dynamics
of the gas are affected by the low-mass primary solids subdisc.

\begin{table*}
 \centering
 \begin{minipage}{140mm}
  \caption{Parameters and corresponding Jeans lengths for computational experiments.}
  \begin{tabular}{@{}lllll@{}}
  \hline
   Experiment No     & 1 &  2 &  3 &  4 \\
 \hline
 Mass of gas & $0.52M_0$ & $0.495M_0$ & $0.52M_0$ & $0.52M_0$  \\
 Mass of solids & $0.03M_0$ & $0.055M_0$ & $0.03M_0$ & $0.03M_0$  \\
 Initial gas temperature at $R = 10$ au (K) & 610 & 1335 & 610 & 610  \\
 Initial velocity dispersion of solids (m s$^{-1}$) & 1900 & 285 & 475 & 95  \\
 Initial Jeans length in gas  &  0.42 &   0.97 &   0.42  &  0.42 \\
 Initial Jeans length in solids at $R = 1$ au & 6.45 &  0.2 & 0.4 & 0.016 \\
 Outcome of instability development & 3 spiral  & 10 spiral arms & 5 spiral &
Gas-solid clumps \\
  & arms & in solids  & arms &
in the \\
& in gas  & transformed into & in gas &
inflection points \\
& and  &  3 spiral arms & and &
 of 5 spiral \\
& solids & in gas and solids.  & solids & arms\\
&  & Thermalization of solids  &  & \\

\hline
\end{tabular}
\end{minipage}
\end{table*}

The last mode is the most interesting; we will demonstrate that
the 'two-phase' system (gas containing a subdisc that has low
solid velocity dispersions) can act as a source of the instability
necessary for the formation of self-gravitating gas-solid clumps
in the disc.

In order to demonstrate that all three modes exist in massive
protoplanetary discs, we reproduced the dynamics of some rotations
of a disc with mass $M=0.55M_0$ and radius $R=2R_0=20$ au,
rotating around a central body with mass $M_c=0.45M_0$. It should
be noted that when going from a 3D disc to its quasi-3D
approximation, we retained a fixed ratio between the masses of the
central body and the disc, $M_c/M$. All calculations reported in
this paper used $\gamma^*=5/3$. The ratio of the surface densities
of the primary solids subdisc and the gas disc
($\sigma_{par}/\sigma_{gas}$) was varied in the range of 0.01 -
0.1. This differed from the ratio between the solid and gas phases
in molecular clouds (which has a value between 0.01 and 0.02), due
to the accumulation of the solid phase in the disc, and the
greater thickness of the gas subdisc compared with the solids
subdisc. In the quasi-3D model,
$\sigma_{par}/\sigma_{gas}=m_{par}/m_{gas}$. By specifying the
ratio of the surface densities and the mass of the entire disc, we
determined the masses of the gas and the primary solids subdisc.
Thus, in calculations 1, 3, 4 (Table 1), the mass of gas in the
disc was $M_{gas}=0.52M_0$, and the mass of the particles was
$M_{par}=0.03M_0$, which corresponds to the surface density ratio
$\sigma_{par}/\sigma_{gas} \approx 0.058$. In contrast, in
calculation 2, $M_{gas}=0.495M_0$ and $M_{par}=0.055M_0$, which
gave a surface density ratio of $\sigma_{par}/\sigma_{gas} \approx
0.11$.

We specified the density and temperature (dispersion) of the disc
to give an initial state of unstable equilibrium. Such a choice of
initial state made it possible to follow the main stages of the
development of instability in the calculations. The initial gas
temperature and solid velocity dispersion conditions were set to
provide a certain level of instability in each component of the
disc. Table 1 provides a summary of the parameters and
corresponding Jeans lengths for each run. Fig. 2 (which shows the
initial radial distributions of the temperature and angular
velocity in the gas, for experiment 1) shows that ultrasonic gas
flow with differential rotation was reproduced.

\begin{figure}
\includegraphics[width=84mm]{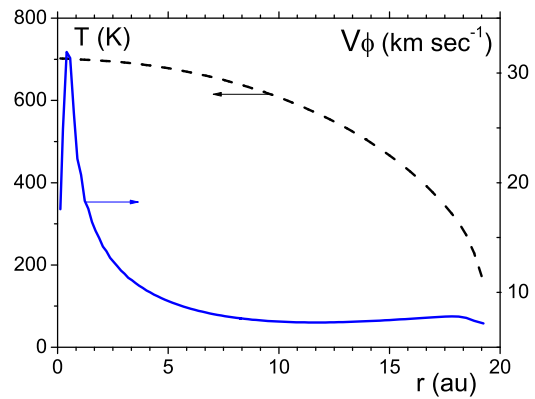}
 \caption{Gas temperature and angular velocity versus radius at initial time, for experiments 1, 3, and 4}
 \label{TandV}
\end{figure}

Experiment 1 was run in a mode where the formation of the disc
structure was determined by the gas, while the gravitational field
of the massive component meant that the solids were also involved
in the structures. In this run, the initial Jeans length $\Lambda$
for the gas (which was constant over the entire disc) was equal to
0.42, which corresponded to a gas temperature profile $T^* \sim
\sigma_{gas}$. For such a dependence, at a radius $R=10$ au the
temperature was 610 K; at a radius $R=1$ au, the temperature was
700 K. The Jeans length in the primary solids subdisc increased
from the centre to the periphery, and reached 6.45 at $R=1$ au,
due to an initial primary solids velocity dispersion of $v_d =
1900$ m s$^{-1}$.

Values for the Toomre parameter $Q=\displaystyle\frac{c_s
\kappa}{\pi G \sigma}$ (which characterises the level of
gravitational stability of a rotating flat disc, and where $\kappa
\approx \Omega$ for discs with near-Keplerian rotation) are shown
in Fig. 3 for calculations 1, 3 and 4, for gases and primary
solids. At a specified gas mass and temperature, a value of $Q <
1$ corresponded to the disc region $R > 2$ au, which indicated
that the radial-ring instability could develop in the gaseous
disc.

\begin{figure}
\includegraphics[width=84mm]{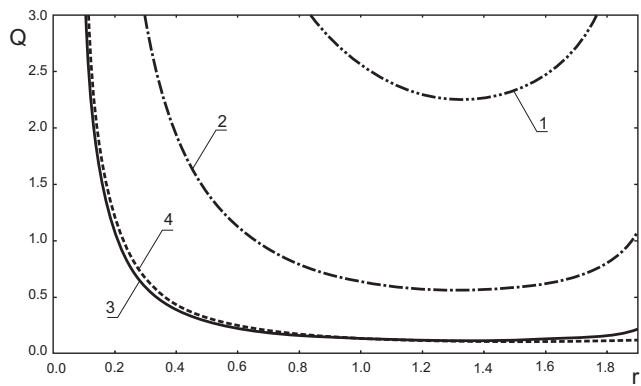}
 \caption{Initial value of the Toomre Q parameter versus radius, for experiments 1, 3, and 4. 1, 2, and 3 show curves for the primary solids subdisc: 1 - experiment 1 (corresponding to an initial thermal velocity $v_d = 1900$ m s$^{-1}$), 2 - experiment 3 ($v_d = 475$ m s$^{-1}$), 3 - experiment 4 ($v_d = 95$ m s$^{-1}$). 4 - gas curve corresponding to all three experiments.}
 \label{Qfig}
\end{figure}

As shown in Fig. 4 in the images from Run 1 (the images show the
surface densities of the gas and particles), broad density rings
had formed in the gaseous disc and on its periphery by the time $t
= 10$. The density distribution of solids was similar to that in
the gas, except for the narrow solid density rings at a radius of
$\approx 8$ au. By the time $t = 15$, azimuthal instability had
developed, resulting in the formation of a three-arm spiral
structure in the gas. The width of the solid spirals was much
smaller than the width of the gas spirals. The solids concentrated
in the spiral gas arms formed their own structure, which was
thinner than the gas structure. At a later time of $t = 23$, this
relatively thin solid structure disappeared, leaving a well
developed three-arm structure; in this case, the dynamics of the
disc were determined by the gas. Therefore, for high initial
velocity dispersions, the bodies showed neither large
inhomogeneities in their density nor any significant
thermalisation of velocities, leading to a smooth spiral structure
that covered the solids; this was the result when the disc
dynamics were determined almost completely by the massive gas
component.

\begin{figure*}
  \includegraphics[width=150mm]{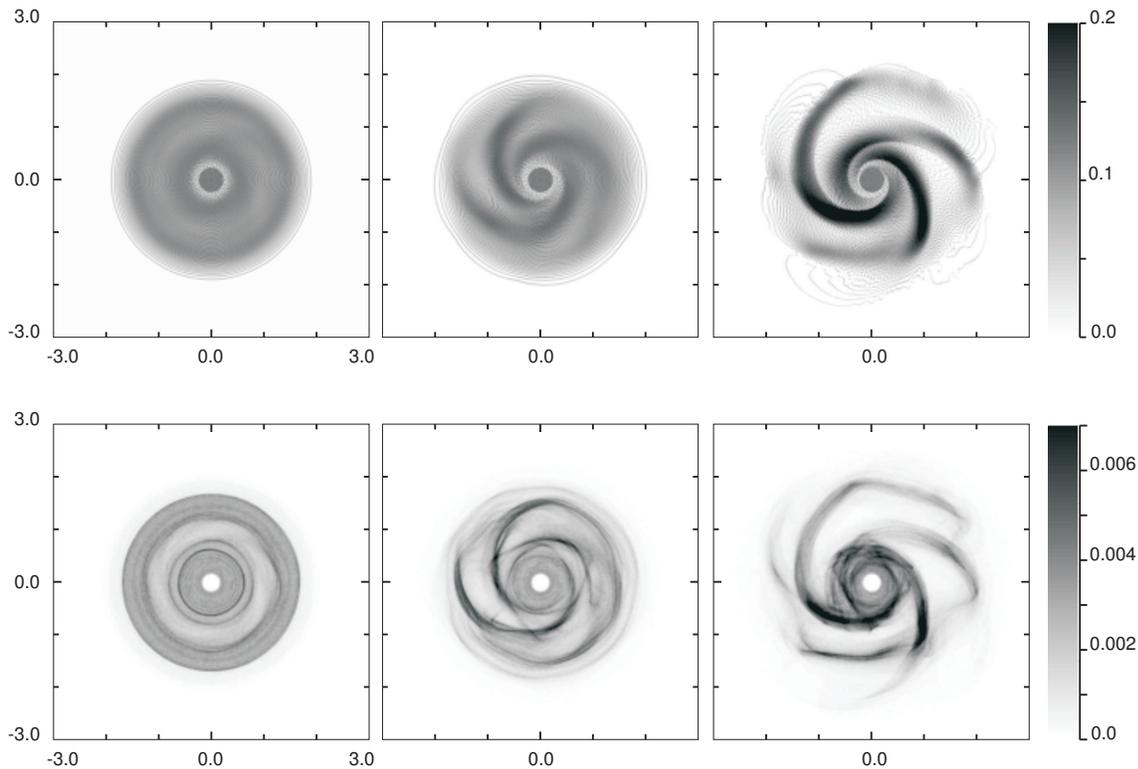}
  \caption{: The formation of spiral arms due to the development of gravitational instability in the gas. The surface density of the gas (top) and the primary solids subdisc in experiment 1, for time points T = 10, 15, and 23. }
\end{figure*}

The benchmark Run 2 --- where the gas temperature increased more
than twofold, and the solid velocity dispersion decreased by a
factor of almost 7 compared with Run 1 --- showed radically
different disc dynamics. This run corresponded to the second case
considered above for the interaction between the particles and
gas. For such disc parameters, the effective Jeans length
$\Lambda$ was determined by the primary solids subdisc, as
illustrated in Fig. 5. By $t = 10$, a strong azimuthal instability
had developed in the low-mass component of the disc, resulting in
the formation of numerous spiral arms. Radial-ring structures in
the solid disc were revealed in the gas. By $t = 20$, further
interactions between the spiral waves in the solid phase had
produced a distinct ten-arm structure in the outer part of the
subdisc, and a complex wave structure in the interior of the
subdisc. Such structures were generally absent in the gas
component. The function $\triangle N=\displaystyle\frac{\partial
N}{\partial v_r} \triangle v_r,$ where $N$-the velocity
distribution function of solids, (see Fig. 6) remained almost
unchanged for $t = 20$. By $t = 30$ (which roughly corresponded to
the next disc rotation over the periphery), the solids were
scattered in patterns determined by their density waves, with
their distribution function showing an increase of nearly an order
of magnitude. Such a distribution function corresponded to a 3- to
4-arm density of solids in the disc (see Fig. 5) and weakly
revealed structures in the gas. The number of particles with
significantly lower velocities (obeying the distribution function
$dv < -0.4$) was still increasing, as was the number of solids
with slightly higher velocities (with $dv < 0.2$); by $t = 50$,
the disc composed of both solids and gas attained an equilibrium
thermalised state. Thus, for this run, the thermalisation time of
the disc (whose instability was higher in the solid component than
in the gas) was smaller than the time taken for the solid subdisc
to decay into separate clumps, under Jeans instability working
against a moving but uninvolved gas.

\begin{figure*}
\includegraphics[width=130mm]{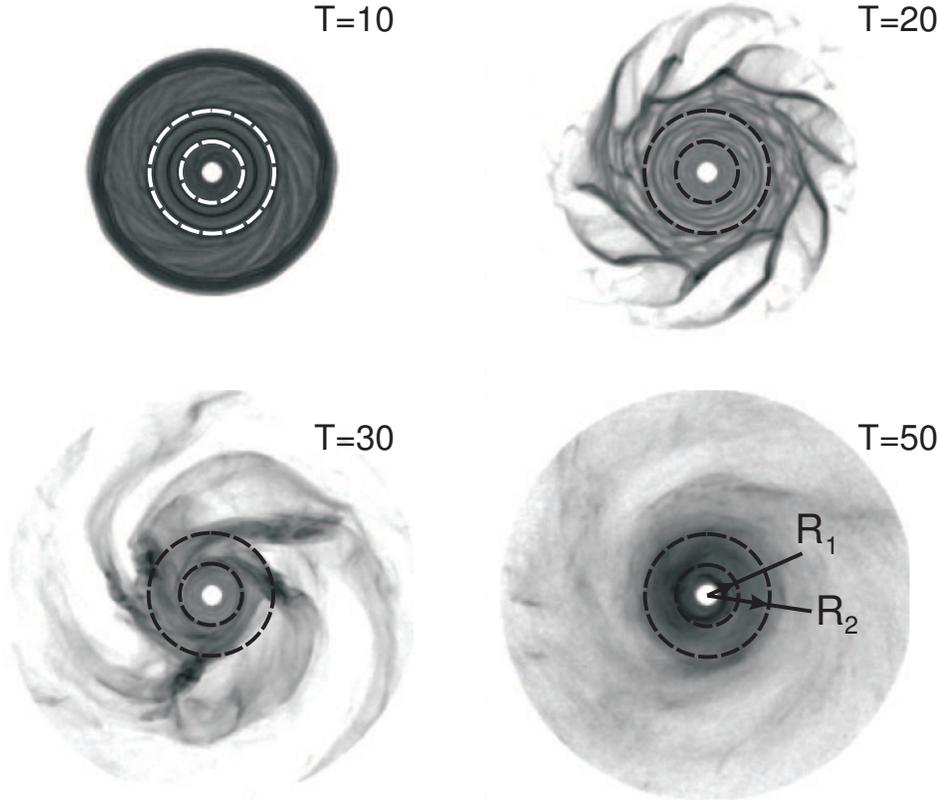}
 \caption{The formation of structures and the development of gravitational instability in a mode where the value of the Toomre parameter for the primary solids subdisc was lower than that for the gas. Logarithm of the primary solids subdisc surface density, for time points T = 10, 20, 30, and 50, for experiment 2.}
 \label{Therm}
\end{figure*}

\begin{figure}
\includegraphics[width=84mm]{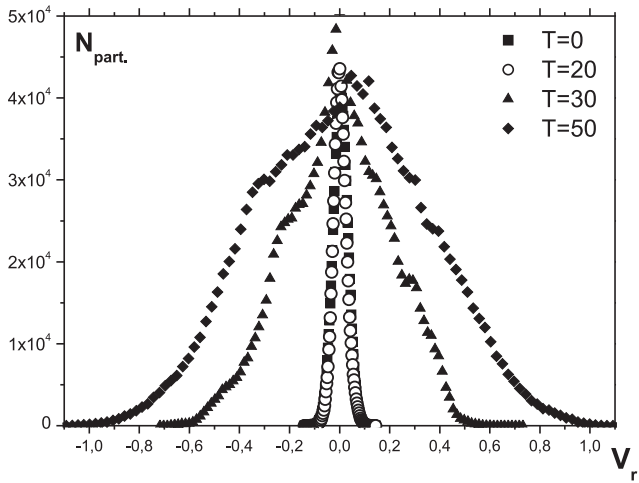}
 \caption {Function $\triangle N=\displaystyle\frac{\partial N}{\partial v_r} \triangle v_r,$ where N-the velocity distribution function of PIC particles in a solid subdisc residing in a ring [r1, r2] = [0.45, 0.9], for time points $T = 0 \ \ \triangle v_r=0.0033,$ $T=20 \ \ \triangle v_r=0.007,$ $T=30 \ \ \triangle v_r=0.011,$ $T=50 \ \ \triangle v_r=0.013,$ in experiment
 2.)}
 \label{Function}
\end{figure}

Experiments 3 and 4 showed the development of two-phase
gravitational instability, with both the massive gas component of
the disc and the primary solids subdisc influencing the formation
of structures. The disc dynamics calculations were performed using
gas parameters similar to those used in experiment 1; the only
change was that the primary solids velocity dispersion was
decreased to $475$ m s$^{-1}$ in experiment 3, and $95$ m s$^{-1}$
in experiment 4.

In Fig. 7, for Run 3, the density distribution in the gas-solid
subdisc indicated that the development of this process in a gas
with a colder low-mass component produced a 5-arm structure
instead of a 3-arm one, after the same amount of time. The lateral
linear dimensions of the individual density waves in the solids
spiral (i.e., the thickness of these entities) were much smaller
than those obtained in experiment 1 for the spiral waves formed
mainly by the gas component. Such a decrease in the thickness of
the solid 'filaments' also took place in the radial-ring structure
on the disc periphery, which was less clear for the gas. The
peripheral structure of the solids was characterised by a strong
interaction with the spiral waves in the middle of the disc, which
led to the breakdown of the azimuthal symmetry. This interaction
produced high-density zones, due to the coupling between the
particles and gas in the radial-ring and spiral-wave domains. This
result raises the possibility that self-gravitation sufficient to
oppose thermalisation factors in the two-phase disc may be
triggered in such zones.

\begin{figure*}
  \includegraphics[width=150mm] {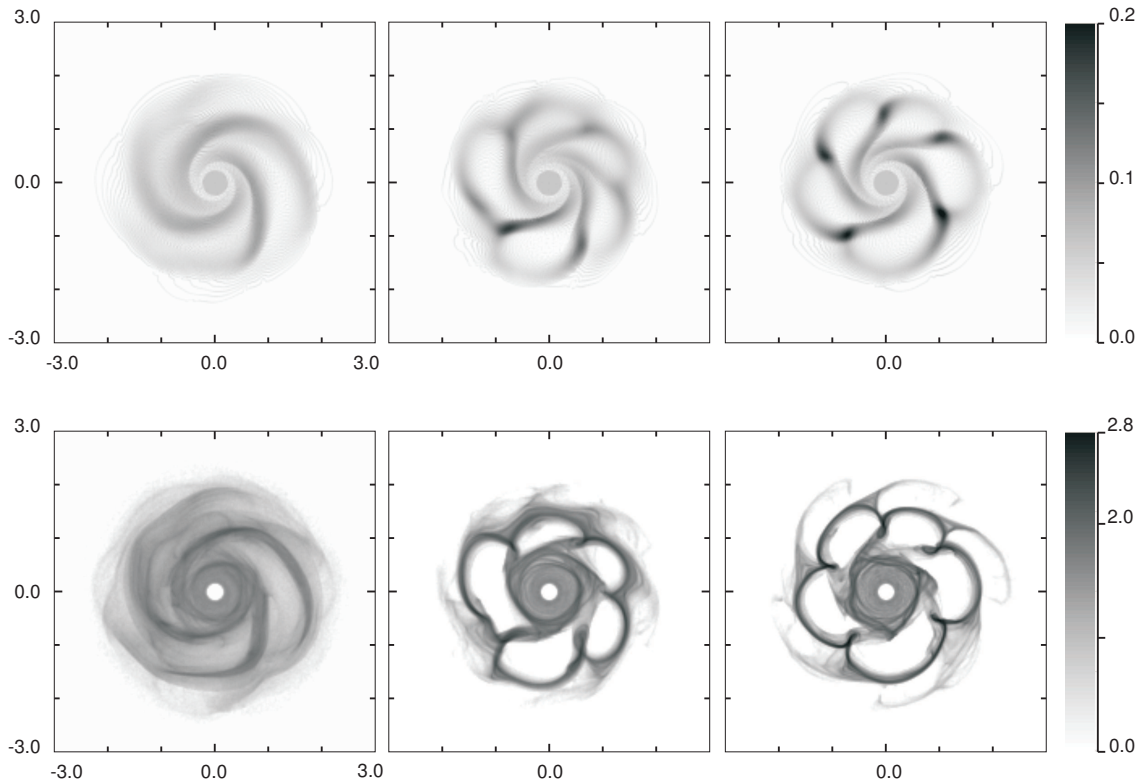}
  \caption{The surface density of gas (top) and the logarithm of the surface density of the primary solids subdisc, for experiments 1, 3, and 4, for time points T = 20, 19, and 22, respectively.}
\end{figure*}

In order to assess this possibility, let us consider the results
of experiment 4, where the solid velocity dispersion was decreased
again by a factor of 5, to approximately 0.1 $c_s$. In this case,
the spiral structure did not change, and the spiral retained its
full number of arms. This confirmed the importance of the role of
the massive and unstable gas disc in the appearance of the spiral
structure. However, over a narrow range, the number of arms
depended on the solid subdisc parameters. The thickness of the
solid-phase arms decreased significantly in comparison with
experiment 3. In the radial-ring and spiral-wave coupling region,
more clumps were formed with high densities of both solid and gas.
Here, the solid phase density exceeded the background values by a
factor of at least ten. The clumps formed in the regions of
minimum Jeans length.
Thus, the development of instabilities in the two-phase disc
resulted in the formation of five areas of increased gas-solid
density, with a possible subsequent local gravitational collapse
of gas and solids. There was insufficient time for the scattering
of solids by density waves to occur in these regions; the time was
also not sufficient for the primary solids subdisc to experience
complete thermalisation, or to pass out of its quasi-equilibrium
state. This conclusion neglects the drag force between gas and
solids, an assumption that is valid for some tens of rotations. At
long times, drag due to the gas should be taken into account, as
it can decrease the velocity dispersion of solids at a certain
radius from the central body.

\begin{figure*}
  \includegraphics[width=150mm]{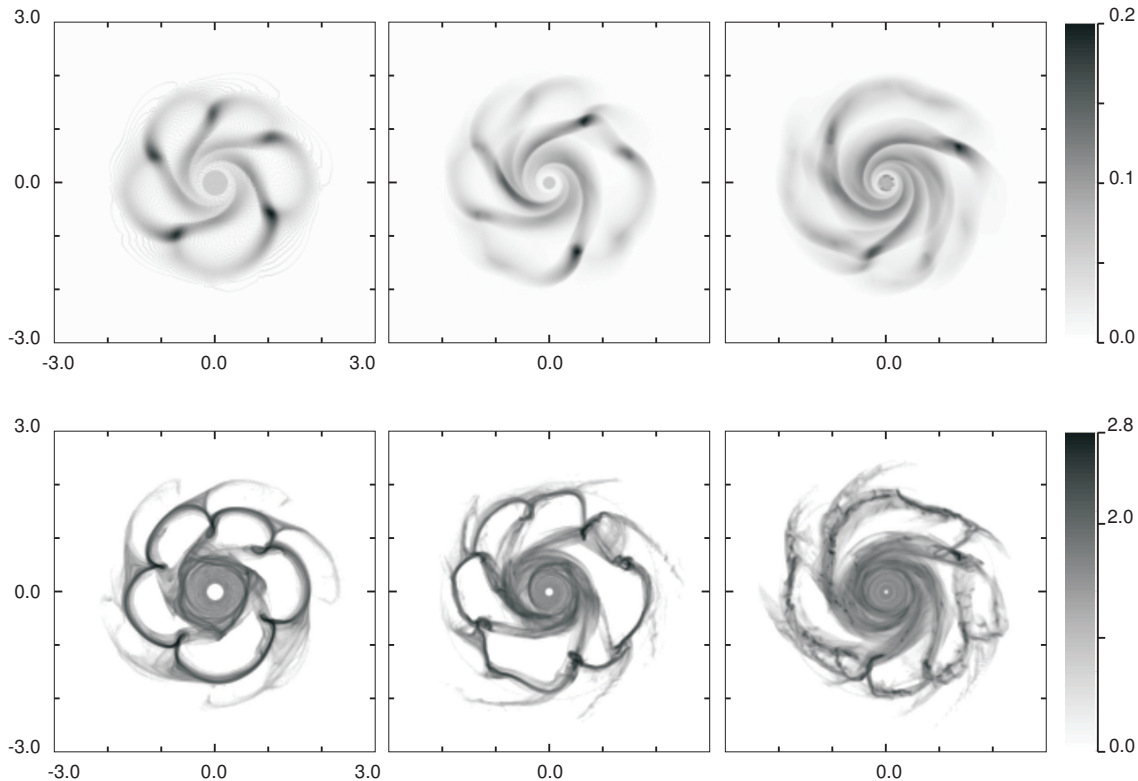}
  \caption{The surface density of gas (top) and the logarithm of the surface density of the primary solids subdisc, for experiment 4 with varying numerical resolution, for time points T = 22, 23, and 23, respectively. The left column - 40000 SPH particles, 2500000 PIC particles, $100\times 128 \times 100$ grid cells, the middle column - 160000 SPH particles, 10000000 PIC particles, $200\times 256 \times 200$ grid cells, the right column - 640000 SPH particles, 40000000 PIC particles, $400\times 512 \times 400$ grid cells.}
\end{figure*}

To make sure that fragmentation in experiment 4 is not a numerical
artefact we calculated this regime with increasing numerical
resolution. We consequently increased the number of SPH and PIC
particles four times and the number of meshpoints twice in each
direction. Fig. 8 shows the surface density of gas and solid
bodies in linear and logarithmic scale respectively. The spiral
structure is formed almost simultaneously in calculations with
increasing numerical resolution. We calculated the development of
physical instability. Generally it shows a discontinuous
dependency of solution on problem parameters (Vshivkov, Nikitin \&
Snytnikov, 2003). So details in density plots are changed with the
numerical resolution variation, when the picture almost remains.
All plots demonstrate the appearance of gas clumps bound with
solid clumps. Increasing the number of PIC and SPH particles and
meshpoints allowed us to reproduce the wave interaction of the
structures with the groups of particles.

These results can be used to explain the mechanism by which gas
giants are formed. The formation of gas clusters due to
gravitational instability has been considered to be a possible
mechanism for the formation of Jupiter and Saturn (Boss 2000).
However, the separate fragmentation of the gas disc was feasible
(in computational experiments) only under strong cooling of the
system during several rotations (Boss 2000, Durisen et al. 2007,
Rice et al. 2006, Meru \& Bate 2010). Our calculations
demonstrated that in a two-phase gravitating medium, the low-mass
part of the disc (composed of metre-sized solids) could initiate
and accelerate the gravitational growth of waves in both
components of the system, for the case of low solid velocity
dispersions.

\section[]{Conclusions}

In this study, we examined a major bottleneck in the understanding
of the process of planet formation; namely, the formation of large
bodies (planetesimals and planet embryos) from metre-sized
boulders in the circumstellar disc. A planetesimal in its emergent
stage is considered as a clump of gas and solids whose
gravitational field preserves its mass when the clump moves. We
have proposed a mechanism that can explain the formation of
planetesimals in massive accreting discs: in this emergent stage,
the mass of a protostar is almost equal to the mass of the
circumstellar disc. Gas leaves the solid subdisc, being reflected
from the equatorial plane, and is scattered to outer space. The
gas temperature drops during this process, under conditions of low
radiation from the low-mass protostar. Solid bodies grow to metric
size with the occurrence of collisions, and lose relative velocity
due to gas drag. In general, the solid component stays in the
equatorial plane of the disc, while the gas leaves it. The
fraction of the solid component therefore increases with respect
to the gas. The increasing solid density, in combination with the
decreasing gas temperature, causes the two-phase disc to transfer
into a marginal state, allowing the development of gravitational
instability of some type.

Our computer simulation showed that self-gravitating clumps were
formed in a massive disc via the development of a 'two-phase'
Jeans instability in the gas-primary bodies medium. These bodies
were of larger than metre-sized, and rotated around the protostar
without the occurrence of collisions per orbital time. In these
unstable conditions, the overall gas dynamics were affected by the
primary solids subdisc, via its gravitational field. This implied
that the possibility of clump formation was determined both by the
rate of gas cooling and its density redistribution, and by the
rate of concentration of large (over 1 m) primary solids, and the
decrease in their velocity dispersions (cooling of primary
solids). We found that a velocity dispersion of 0.1 $c_s$ in the
boulder subdisc was sufficient to cause the formation of
self-gravitating clumps of gas and boulders. In such regimes, the
time taken for the formation of collapsing objects was less than
time taken for boulders to dissipate in the density waves of the
medium.

\section*{Acknowledgments}

We would like to thank the anonymous referee, whose constructive comments led to an improvement of the paper. Our work was supported by the RAS Presidium programs 'Biosphere
origin and evolution' and 'Origin, structure and evolution of
objects in the Universe', as well as SB RAS Integration Project
No. 26 'Mathematical models, numerical methods and parallel
algorithms for solving big problems of SB RAS and their
implementation on multiprocessor supercomputers', and Russian
Federation President Grant for the Leading Scientific Schools NSh
3156.2010.3.

\bsp

\label{lastpage}

\end{document}